# Electronic origin of superabundant vacancies in Pd hydride under high hydrogen pressures

Degtyareva V.F.

Institute of Solid State Physics Russian Academy of Sciences,
Chernogolovka, 142432 Russia

Fax: +7(496) 524 9701     E-mail: degtyar@issp.ac.ru

## Introduction

Formation of a defect fcc structure containing metal vacancies in Ni-H and Pd-H alloys was observed experimentally at high temperature annealing under high hydrogen pressure [1]. This phenomenon called superabundant vacancy (SAV) formation has been substantiated by the observation of lattice contraction and by density reduction for some metal-hydride systems.

Variation of lattice parameter of Pd hydride with time, measured under a hydrogen pressure of 5 GPa is shown in Fig.1. The gradual lattice contraction, which occurred at 800 $^{o}$C, is caused by metal vacancies in the fcc lattice approaching to the ultimate structure $M_3VacH_4$. The SAV formation is provided by the formation of a vacancy-ordered structure in the fcc hydrides ($Ni_3VacH_4$, $Pd_3VacH_4$), which is likely to be a $L1_2$-type structure ($Cu_3Au$ type metal lattice).

For the mechanism of SAV formation, it was inferred that vacancy - hydrogen (Vac-H) binding energy should be responsible. The physical understanding of SAV formation is the lowering of the energy of interstitial H atoms neighboring metal vacancies by formation of the Vac-H clusters.

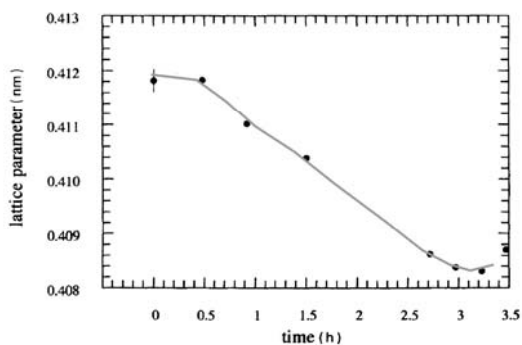

Fig.1. Temporal variation of the lattice parameter of Pd hydride at 800 $^{o}$C and a hydrogen pressure of 5 GPa (after Fukai [1]).

## Theoretical model and consideration

Here we consider the problem of vacancy formation in metal hydrides in relation to the electronic energy as one of the main contribution to the crystal energy. Significance of this part of energy is essential for so-called electronic compounds or Hume-Rothery phases. Classical example of an alloy system with Hume-Rothery phases is the Cu – Zn system with a phase sequence

fcc – bcc – complex cubic – hcp.

Phase boundaries are correlated with a number of the valence electrons per atom or electron concentration, e/a.

A theoretical explanation of phase stability in Cu – Zn and related systems was suggested by Mott and Jones [3] within a nearly-free-electron model considering configurations of Brillouin zone and Fermi surface. Assuming the Fermi surface like a sphere, the Fermi radius is defined as $k_F = (3\pi z/V)^{1/3}$, where z is the number of valence electrons per atom and V is atomic volume. Electron energy decreases if a Fermi level is close to a Brillouin zone plane due to formation of the energy gap resulting in the increase of density of states and lowering the Fermi level. The overall crystal energy reduces owing to the decrease of the band structure energy.

The necessary condition is closeness of Fermi radius to a Brillouin zone plane defined by a half of the reciprocal lattice vector $q_{hkl}$: $k_F = ½ q_{hkl}$. For the fcc structure one should take a (111) plane resulting in the contact condition at z =1.36 electron/atom, e/a. For the bcc structure taking a (110) plane results in z = 1.5 e/a. These values of electron concentration define the phase boundaries of fcc and bcc phases in many binary alloy systems of Hume-Rothery type.

In certain alloy phases it has been found that the number of atoms per unit cell decreases as the phase became richer in the component of higher valency. Vacancy defects occur to maintain an optimum electronic energy. Equiatomic Ni – Al alloy is a typical example of a defect phase [2]. At 50 % it possesses CsCl, bcc related structure. With an increase of Al content, lattice parameter and density decrease which is caused by the formation of vacancies in Ni sublattice. As shown in Fig. 2 the lattice parameter falls down below 49 at.% Ni and density curve breaks away from the calculated curve (dotted line) and drops well below it. For alloy with ~45 at.% Ni estimations from density measurements give the number of atoms per unit cell 1.84 instead of 2, however the total number of valence electrons per cell is keeping equal to 3 to satisfy the Hume-Rothery condition.

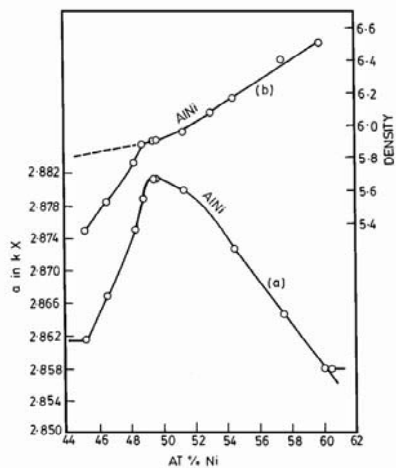

Fig.2. Lattice parameters and density of the AlNi bcc phase as a function of composition (after Pearson [2]).

The Hume-Rothery model can be extended to understanding of defect structures containing metal vacancies in metal-hydride systems, as in Pd – H. It is necessary to assume that atom H inserted in the metal lattice loses his electron which enters into a join valence band of a metal hydride. This additional electron is distributed between d and s levels in transition metal hydrides as was shown by investigations of magnetic properties in some hydrides by Antonov [4].

For Pd and Ni hydrides by estimation of the number of valence electrons per metal atom one should assume that d-band of these metals is filled and additional electrons inserted by H atoms are moved into the s-band. This assumption results for $Pd_3VacH_4$ in the number of valence electrons per metal atom defined as $4/3 = 1.33$ e/a. This value is very close to the limiting condition for the fcc phase in the Hume-Rothery alloys $z = 1.36$ e/a.

Fig.3 shows the configuration of Brillouin zone for the fcc lattice and the Fermi sphere in hydrides PdH and $Pd_3VacH_4$. Fermi sphere radius is defined for values of z equal to 1 and 1.33, respectively.

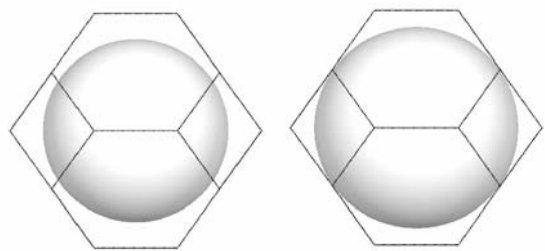

Fig.3. Configuration of Brillouin zone and Fermi surface (sphere) for fcc lattice in hydrides PdH (left) and $Pd_3VacH_4$ (right).

Considerations following the Mott and Jones approach suggest that the closeness of Fermi sphere to a Brillouin zone plane and formation of an energy gap result in increase of density of states and lowering of the electron energy. One can imagine that by subjecting hydride at such condition as high temperature and high hydrogen pressure diffusion processes are accelerated for metal atoms, leading to formation of a defect structure, if such a structure has a minimal energy. These conditions allow the formation of vacancy-ordered structure similar to the ordered alloy phase of $Cu_3Au$ type.

**Summary**
Thus, formation of vacancies in the fcc structure of Pd hydride and several other metal hydrides can be accounted for by electronic origin assuming that valence electron energy is minimized due to Hume-Rothery effects.